\documentclass[conference]{IEEEtran}
\usepackage{cite}
\usepackage{amsmath,amssymb,amsfonts}
\usepackage{algorithmic}
\usepackage{graphicx}
\usepackage{textcomp}
\usepackage{xcolor}
\usepackage{listings}
\def\BibTeX{{\rm B\kern-.05em{\sc i\kern-.025em b}\kern-.08em
    T\kern-.1667em\lower.7ex\hbox{E}\kern-.125emX}}

\makeatletter
\@ifundefined{showcaptionsetup}{}{
 \PassOptionsToPackage{caption=false}{subfig}}
\usepackage{subfig}
\makeatother

\usepackage{eso-pic}
\newcommand\AtPageUpperMyright[1]{\AtPageUpperLeft{
 \put(\LenToUnit{0.5\paperwidth},\LenToUnit{-1cm}){
     \parbox{0.5\textwidth}{\raggedleft\fontsize{9}{11}\selectfont #1}}
 }}
\newcommand{\conf}[1]{
\AddToShipoutPictureBG*{
\AtPageUpperMyright{#1}
}
}

\usepackage{tikz}
\usepackage{textcomp}
\newcommand\copyrighttext{%
  \footnotesize \textcopyright 2019 IEEE.  Personal use of this material is permitted.  Permission from IEEE must be obtained for all other uses, in any current or future media, including reprinting/republishing this material for advertising or promotional purposes, creating new collective works, for resale or redistribution to servers or lists, or reuse of any copyrighted component of this work in other works.
  DOI: 10.1109/IPDPSW.2019.00047}
\newcommand\copyrightnotice{%
\begin{tikzpicture}[remember picture,overlay]
\node[anchor=south,yshift=10pt] at (current page.south) {\fbox{\parbox{\dimexpr\textwidth-\fboxsep-\fboxrule\relax}{\copyrighttext}}};
\end{tikzpicture}%
}
 
\title{Delta-stepping SSSP: from Vertices and Edges to GraphBLAS Implementations}
\author{
\IEEEauthorblockN{Upasana Sridhar, Mark Blanco, Rahul Mayuranath, \\
Daniele G. Spampinato, Tze Meng Low}
\IEEEauthorblockA{\textit{Department of Electrical and Computer Engineering} \\
\textit{Carnegie Mellon University}\\
Pittsburgh, PA, USA\\
\{upasanas, markb1, frahulma\}@andrew.cmu.edu, \{spampinato, lowt\}@cmu.edu}
\and
\IEEEauthorblockN{Scott McMillan}
\IEEEauthorblockA{\textit{Software Engineering Institute} \\
\textit{Carnegie Mellon University}\\
Pittsburgh, PA, USA\\
smcmillan@sei.cmu.edu}
}

\begin{document}

\maketitle
\conf{2019 IEEE International Parallel and Distributed Processing Symposium Workshops (IPDPSW)}
\copyrightnotice

\begin{abstract}
GraphBLAS is an interface for implementing graph algorithms. Algorithms implemented using the GraphBLAS interface are cast in terms of linear algebra-like operations. However, many graph algorithms are canonically described in terms of operations on vertices and/or edges. Despite the known duality between these two representations, the differences in the way algorithms are described using the two approaches can pose considerable difficulties in the adoption of the GraphBLAS as standard interface for development.
This paper investigates a systematic approach for translating a graph algorithm described in the canonical vertex and edge representation into an implementation that leverages the GraphBLAS interface. 
We present a two-step approach to this problem. First, we express common vertex- and edge-centric design patterns using a linear algebraic language. Second, we map this intermediate representation to the GraphBLAS interface. We illustrate our approach by translating the delta-stepping single source shortest path algorithm from its canonical description to a GraphBLAS implementation, and highlight lessons learned when implementing using GraphBLAS.
\end{abstract}


\section{Introduction}
A graph $\mathcal G = (V, E)$ is traditionally defined as a collection of vertices $(V)$ and edges $(E)$ that represent relationships between pairs of vertices. As such, many graph algorithms are described in terms of operations on the vertices and/or edges. Algorithms that can be described as independent operations over the vertices are often referred to as vertex-centric algorithms, while edge-centric algorithms are algorithms cast in terms of operations on the edges.  
Several graph-processing frameworks~\cite{ligra, pregel, kyrola2012graphchi, roy2013x, voegele2017parallel} have been designed to support either vertex- or edge-centric algorithms.

The GraphBLAS~\cite{graphblasapi,kepner2016mathematical} is a community effort to provide a standardized application programming interface (API) based on the language of linear algebra for the implementation of graph algorithms. This standardized interface allows for a separation of concerns between algorithm writers and library developers. By expressing algorithms in terms of GraphBLAS building blocks, algorithm developers can focus on the use and development of graph algorithms without worrying about performance-driven details of the implementation. Library developers also benefit from a standardized interface as they can focus on delivering performance for a specific set of target functionalities. In addition, the use of linear algebra-like operations allows library developers to leverage decades of expertise in high-performance sparse linear algebra to provide high-performance parallel implementations to the graph community.

However, the adoption of a matrix representation-based standard API
faces considerable challenges considering that algorithm writers are prone to describing their algorithms in terms of operations on vertices and edges despite the equivalence of the two representations~\cite{APP}.
This choice is often made based on the notion that adjacency matrices necessarily
lead to a $O(N^2)$ space requirement~\cite{Latapy:2008:MTC:1450344.1450496}. On the other hand, while the GraphBLAS interface describes the operations using matrices and vectors~\cite{kepner2011graph}, their implementation takes advantage of the sparse nature of these objects. 
This representational dilemma presents the algorithm developer with a crucial question: 
how should an algorithm expressed using vertices and edges be expressed using linear algebra and implemented in an API such as the GraphBLAS? While several graph algorithms have been 
studied and implemented using the language of linear algebra~\cite{kepner2011graph}, many others remain unimplemented. Without a systematic approach to the translation problem, it is unclear if this lack of algorithmic implementation is a matter of effort or if there are fundamental restrictions with the implementation API.

This paper represents a \textit{first} attempt at the definition of a systematic methodology for translating graph algorithms expressed using operations on vertices and edges into a linear algebraic formulation implementable using the GraphBLAS API. The approach we propose operates in two steps: First, we convert the vertex- and edge-based
operations in the input algorithm into their matrix-based counterpart using a predefined set of mappings between commonly-used vertex- and edge-based data access patterns, data structures, and operations and their linear algebraic equivalents. Second, we translate the newly obtained linear algebraic entities into GraphBLAS data structures and function calls.
We demonstrate our approach using the delta-stepping single source shortest path ($\Delta$-stepping) algorithm~\cite{dsssp}, one of the algorithms in the Graph Analysis Benchmark (GAP) suite~\cite{GAP}. 
We also report 
our implementation experience with using the GraphBLAS C and C++ APIs. Finally, we highlight opportunities for improvement in performance and parallelism.



\section{Preliminaries}
A graph can be represented in GraphBLAS using either its adjacency or its incidence matrix. As both representations contain the same information, we restrict ourselves to only uses of adjacency matrices. 
From this, we identify linear algebraic constructs relevant to our approach: matrices, vectors, and operations on these constructs. We use these to formulate operations on sets, vertices, and edges as used in traditional graph algorithms in the language of linear algebra.

\subsection{Adjacency matrices}
An adjacency matrix representation of the graph $\mathcal G = (V, E)$ is a $|V| \times |V|$ matrix $A_{\mathcal G}$.
An edge between $v_i$ and $v_j$ where $v_i, v_j \in V$ is represented as an element in the $(i,j)^{th}$ position of $A_{\mathcal G}$. 

Non-empty elements in the $i^{th}$ row represents outgoing edges from vertex $v_i$, where $v_i \in V$, and the $i^{th}$ column represents the incoming edges into vertex $v_i$. We assume that our graphs are simple and there are no self-loops. This means that the 
diagonal elements of the adjacency matrix $A_{\mathcal G}$ are empty.

Notice that storing the graph as an adjacency matrix implicitly provides a unique identifier for each vertex. This identifier can be used to index into particular rows and columns of the adjacency matrix that correspond to specific vertices. Note also that assigning unique identifiers to vertices is a technique often found in traditional graph algorithms to avoid overcounting.

\subsection{Operation on the vertices}
Algorithms that solely perform operations from the viewpoint of a vertex are vertex-centric algorithms and are often the target of ``think like a vertex" frameworks~\cite{VertexFrameworks}. In particular, algorithms that fit this framework typically perform computation on incoming and outgoing edge values of a particular vertex. In addition, all vertices in the graph are processed independently and possibly in parallel.

Recall that incoming edge values to a vertex $v$ are captured as elements in the $v^{th}$ column of the adjacency matrix. Partitioning the adjacency matrix by columns yields the following:
\[
A_{\mathcal G} \rightarrow 
\left(
\begin{array}{c|c|c|c}
     a_0 & a_1 & \hdots & a_{|V|-1}\\
\end{array}
\right).
\]
Operations on the incoming edge values can be cast as operations on the appropriate column vector, i.e.,
\[
\left(
\begin{array}{c|c|c|c}
     f_0(a_0) & f_1(a_1) & \hdots & f_{|V|-1}(a_{|V|-1})\\
\end{array}
\right).
\]

Operations on the out-going edge values can be described using a linear algebraic formulation in a similar manner where the adjacency matrix is first transposed and then the operations are applied.

\subsection{Operation on the edges}
Algorithms whose operations are performed on all edges simultaneously are often known as edge-centric algorithms. Values on edges are captured as the individual entries of the adjacency matrix. This means that an edge-centric algorithm will update or use all values of the matrix simultaneously.

An operation that is applied to all edges can be cast in the language of linear algebra in a number of ways. The first approach is to treat the operation on all edges similar to how a scalar element can be multiplied against all elements in the adjacency matrix, i.e.,
\[
\beta A_{\mathcal G},
\]
where $\beta$ is the operation to be applied to each edge. This is simply a point-wise operation.

The second approach for capturing operation on the edges is for the situation when the edge values in the adjacency matrix are the output of a series of linear algebra operations (e.g. matrix-matrix multiplication) such as in the computation of the graph's K-truss~\cite{ktruss_2018}:
\[
S = A_{\mathcal G}^T A_{\mathcal G}.
\]
In this case, extraneous elements may be created in $S$ that results in fill-in. Under such a scenario, it is important to apply a Hadamard, or element-wise, product of the result with the original $A_{\mathcal G}$ to eliminate these spurious entries, i.e.,
\[
S = A_{\mathcal G}^T A_{\mathcal G} \circ A_{\mathcal G}.
\]


\subsection{Sets}
Sets are often used in graph algorithms to denote groups of vertices or edges. We denote sets of vertices as vectors of size $|V|$, while sets of edges are viewed as matrices of size $|V| \times |V|$. Members of a set are often denoted by a non-zero entry at the location of the matrix or vector set corresponding to the label of the member elements.

\subsection{Filtering}
Often, it is necessary to selectively apply an operation to vertices or edges based on certain criteria. An example of filtering on vertices is the deactivation of a vertex in a vertex-centric algorithm.

One possibility of describing the selection is to extract a subgraph $\mathcal G_1$ by using selector matrices that are multiplied against the adjacency matrix of the original graph $\mathcal G$ from the right and left~\cite{kepner2016mathematical}.

An alternative approach, and the one we select, is to perform a filtering operation where a binary mask $B$ of the same size as our graph is first constructed. The Hadamard product is then performed with the adjacency matrix $A_{\mathcal G}$ to obtain the
filtered adjacency matrix $A_{\mathcal G_1}$. We represent this element-wise product as follows:
\[
\begin{array}{rcl}
A_{\mathcal G_1} &=& B \circ A_{\mathcal G} \\
     &=& A_{\mathcal G} \circ B \\
\end{array}
\]

Filtering operations can also be performed on sets of vertices in the same manner. Instead of a binary matrix $B$, a binary vector $b$ is used instead.

\section{Delta Stepping Single Source Shortest Path}
The goal of all single source shortest path algorithms is to identify the shortest path, defined as the path with minimum total weight, from a source vertex $s$ to each of the vertices $v \in V$, $v \neq s$, from a given graph $\mathcal G = (V, E)$.
The delta-stepping single source shortest path ($\Delta$-stepping) algorithm was introduced by Meyer and Sanders~\cite{dsssp} as an attempt to improve parallelism in Dijkstra's original SSSP algorithm~\cite{dijkstra}. 

The $\Delta$-stepping algorithm is a label-correcting algorithm in which the tentative distances ($tent(v_i)$) from the source vertex $s$ to all other vertices $v_i$ are iteratively updated through the application of edge relaxation operations.

\subsection{Initialization}
Initialization of the $\Delta$-stepping algorithm starts by first splitting the outgoing edges for all vertices $v_i \in V$ into  two edge sets: \emph{light($v_i$)} and \emph{heavy($v_i$)}. An edge $(v_i, v_j)$ that goes from vertex $v_i$ to vertex $v_j$ is
assigned to the light edge set of $v_i$ if the weight on the edge, $\delta(v_i, v_j)$, is less than or equal to a predefined threshold $\Delta$. Otherwise, the edge is assigned to the set \emph{heavy($v_i$)}. In addition, the tentative distance to $v_i$ from $s$ is set to infinity ($\infty$) for all $v_i \in V$. The distance to $s$ from $s$ is 0.

Mathematically, we can describe the initialization as three separate operations that are applied on all vertices as follows:
\[
\begin{array}{rcl}
\mbox{\bf for all } v_i \in V\\
heavy(v_i) &=& \{(v_i, v_j) \in E : \delta(v_i, v_j) > \Delta \} \\
light(v_i) &=& \{(v_i, v_j) \in E : \delta(v_i, v_j) \leq \Delta\} \\
tent(v_i) &=& \infty
\end{array}
\]

\subsection{Defining buckets}
A distinctive feature of the $\Delta$-stepping algorithm is the use of buckets to hold vertices whose tentative distance from the source vertex $s$ is within a specific range. Specifically, the bucket $B_i$, $i \geq 0$ is the set of vertices whose tentative distance from the source is 
between $i\Delta$ and $(i+1)\Delta$, i.e.,
\[
  B_i = \{v \in V : i\Delta \leq tent(v) < (i+1)\Delta \}.
\]
At the start, only the source vertex is assigned to bucket $B_0$ with a distance of 0.

\subsection{Edge relaxation}
The algorithm proceeds by first identifying the smallest non-empty bucket. 
For a vertex $v$ in that bucket, edge relaxation is applied on all edges in \emph{light(v)}. Specifically, edge relaxation performs three tasks:
\begin{enumerate}
    \item {\em Identifies reachable vertices}. Edge relaxation identifies reachable vertices $v_j$ from vertex $v$ where $(v, v_j) \in light(v)$.
    
    \item {\em Computes new tentative distances}. Having identified reachable vertices from $v$, edge relaxation also generates requests to update the tentative distances of the reachable vertices $v_j$ with the new distance $tent(v) + \delta(v, v_j)$. If the current tentative distance $tent(v_j)$ is greater than the requested updated distance, the tentative distance of vertex $v_j$ is updated, i.e.,
    \[
        tent(v_j) = \min(tent(v_j), tent(v) + \delta(v, v_j)).
    \]
    \item {\em Reassigns vertices to buckets}.  As the membership of vertex $v$ to a particular bucket is determined by the tentative distance from $v$ to $s$, updating the tentative distance of $v$ causes a change to the bucket membership of $v$ from 
    \[
        B_{\lfloor tent(v_j)/\Delta \rfloor} \mbox{ to } B_{\lfloor (tent(v) + \delta(v, v_j))/\Delta \rfloor}.
    \]
\end{enumerate}

In practice, Meyer and Sanders~\cite{dsssp} proposed to simultaneously process all light edges from all vertices in the current bucket. The result is a set $Req$ of tuples containing the reachable vertices and their potentially new tentative distances:
\[
Req = \{(v_j, tent(v) + \delta(v, v_j)): v \in B_i \; \wedge 
\;(v, v_j) \in light(v)\}.
\] 
Doing so simultaneously empties the bucket $B_i$.
Next, each of these requests is evaluated to determine if 
tentative distances of reachable vertices need to be updated. If so, the 
membership of the updated vertices are also changed.  This second step is performed by the \emph{relax} operation: 
\[
\small
\begin{array}{l}
\mbox{procedure relax($v$, $new\_dist$)} \\
\quad \mbox{if $new\_dist$} < tent(v) \\
\quad \quad B_{\lfloor tent(v)/\Delta \rfloor} = B_{\lfloor tent(v)/\Delta \rfloor} - \{v\} \\
\quad \quad B_{\lfloor new\_dist/\Delta \rfloor} = B_{\lfloor new\_dist/\Delta \rfloor} \cup \{v\} \\
\quad \quad tent(v) = new\_dist
\end{array}
\]
This process may reintroduce vertices into $B_i$, which would then require
us to reiterate over the new set of light edges. 
This process is repeated until the bucket is empty. At this point in time, heavy edges are not required to be relaxed as vertices reachable by traversing a heavy edge will not result in new vertices being introduced into $B_i$.

\subsection{Tracking processed vertices}
Once the current bucket $B_i$ is emptied, the heavy edges from all vertices processed previously can then be relaxed. This requires us to keep track of the vertices that had been in $B_i$. This is performed by updating a set $S$ with vertices in $B_i$:
\[
S = S \cup B_i.
\]

Heavy edges from vertices in $S$ are relaxed in the same manner as light edges: first a set of requests are generated, and then the \emph{relax} operation is applied to each of the generated requests.

The process is repeated with the next non-empty bucket until there are no more non-empty buckets. The overall $\Delta$-stepping algorithm is summarized on the right side of Fig.~\ref{fig:DSSSP}. 

While we have only described the sequential algorithm, the authors proposed for $\Delta$-stepping to be parallelized by processing all vertices in a processing phase of the current bucket. A processing phase is defined as the simultaneous relaxation of all light or heavy edges. 




\begin{figure*}[t]
    \centering
    \[
\small
\begin{array}{l|l}
\mbox{\bf Linear Algebraic Forumlation} & \mbox{\bf Meyer \& Sanders} \\ 
\\
A, A_H, A_L \in \mathbb R^{|V| \times |V|}&\texttt{procedure relax(v, new\_dist)} \\
s, i \in \mathbb N &\quad \texttt{if new\_dist < tent(v)} \\
\Delta \in \mathbb R &\quad \quad \texttt{$B_{\lfloor \texttt{tent(v)}/\Delta \rfloor} = B_{\lfloor \texttt{tent(v)}/\Delta \rfloor} - \{\texttt{v}\}$} \\
t, t_{Req} \in \mathbb R^{|V|} &\quad \quad B_{\lfloor \texttt{new\_dist}/\Delta \rfloor} = B_{\lfloor \texttt{new\_dist}/\Delta \rfloor} \cup \{\texttt{v}\} \\
t_{B_i}, S \in \mathbb N^{|V|} &\quad \quad \texttt{tent(v) = new\_dist}
\\
\\
A_H = A \circ (A > \Delta) & \mbox{\texttt{heavy(v) = \{(v, w) $\in$ E: c(v, w) $>  \Delta$\}}}\\
A_L = A \circ (0 < A \leq \Delta) & \mbox{\texttt{light(v) = \{(v, w) $\in$ E: c(v, w) $ \leq \Delta$\}}} \\
t = \infty & \mbox{\texttt{tent(v) = $\infty$}}\\              
t[s] = 0\quad  & \mbox{\texttt{relax(s, 0)}}\\              
i = 0 & \mbox{\texttt{i = 0}}\\ 
 \mbox{\bf while } (t \geq i\Delta) \neq 0 \mbox{ \bf do }            & \mbox{\texttt{\textbf{while} $\neg$isEmpty(B) \textbf{do}}}\\
\quad s = 0  & \quad \mbox{\texttt{S = $\emptyset$}} \\
\quad t_{B_i} = (i\Delta \leq t <(i+1)\Delta) & 
\quad \mbox{\texttt{// By def. }} \texttt{B[i] = }\{\texttt{v} \in V : \texttt{i}\Delta \leq \texttt{tent(v)} < \texttt{(i+1)}\Delta \}\\
\quad\mbox{{\bf while} $t_{B_i} \neq 0$ {\bf do}}    & \quad\mbox{\texttt{\textbf{while} $\neg$isEmpty(B[i]) \textbf{do}}}\\
 \quad\quad t_{Req} = A_L^T(t \circ t_{B_i}) &\quad\quad  \mbox{\texttt{Req = \{(w, tent(v) + $\delta$\texttt{(v, w)) : v }$\in$ B[i] $\wedge$ (v, w) $\in$ light(v)\}}} \\
\quad\quad S = ((S + t_{B_i}) > 0), t_{B_i} = 0 & \quad\quad \mbox{\texttt{S = S $\cup$ B[i]; B[i] = $\emptyset$}} \\
\quad\quad t_{B_i} = (i\Delta \leq t_{Req} <(i+1)\Delta) \circ (t_{Req} < t)&  \quad\quad \mbox{\texttt{\textbf{foreach} (v, x) $\in$ Req \textbf{do} relax(v, x) }}\\
\quad\quad t = \min(t, t_{Req})&\\
\quad \mbox{\texttt{\bf od}} & \quad \mbox{\texttt{\bf od}}\\
\quad t_{Req} = A^T_H(t \circ S)& \quad\mbox{\texttt{Req = \{(w, tent(v) + $\delta$\texttt{(v, w)): v }$\in$ S $\wedge$ (v, w) $\in$ heavy(v)\}}}\\

\quad t = \min(t, t_{Req})& \quad\mbox{\texttt{\textbf{foreach} (v, x) $\in$ Req \textbf{do} relax(v, x)}}\\
\quad i = i + 1 & \quad\mbox{\texttt{i = i + 1}}\\
 \mbox{\bf od } &  \mbox{\bf od } \\
\end{array}
\]
    \caption{Left: $\Delta$-stepping single source shortest path in the language of linear algebra. Right: The corresponding description using sets of vertices and edges is placed alongside the linear algebra formulation for ease of comparison. }
    \label{fig:DSSSP}
\end{figure*}

\section{Linear Algebraic Delta Stepping SSSP}\label{sec:ladsssp}
The first step in our approach is to translate the vertex and edge representation of the algorithm into linear algebra-like operations. 

\subsection{Initialization}
The $\Delta$-stepping algorithm starts by splitting outgoing edges for all vertices into two sets, light and heavy, and initializing the tentative distance for the source vertex $s$ to 0 and all other vertices to infinity. 

Initializing the tentative distance for all vertices can be described by defining a vector $t$, such that 
\[
t = \infty \mbox{ and }
t[s] = 0.
\]

To split the edges into two sets, first recall that the $i^{th}$ row of the adjacency matrix represents outgoing edges from vertex $v_i$. The light edges for $v_i$ are the edges in the $i^{th}$ row whose weight is less than or equal to the threshold $\Delta$. Since we have to identify light edges for all vertices, this means that for all rows in the adjacency matrix, we have to apply an element-wise check for light edges. This check of the entire adjacency matrix $A$ can be described as
\[
0 < A \leq \Delta.
\]
Notice that applying the check to all elements yields a binary matrix instead of a matrix that contains only the light edges and their weights. It is important to note that the locations of the 1s in the resulting binary matrix are also the exact locations of edges that are light edges. Therefore, to obtain the desired matrix, we can perform a Hadamard product of this binary matrix against the original matrix $A$ to form $A_L$, i.e.,
\[
A_L = A \circ (0 < A \leq \Delta),
\]
to obtain the matrix with only light edges for all vertices. 

The heavy edges can be separated similarly into a matrix $A_H$ with
\[
A_H = A \circ (A > \Delta).
\]

\subsection{Defining Buckets}
Recall that the definition of the buckets is given by
\[
B_i = \{v \in V : i\Delta \leq tent(v) < (i+1)\Delta \}.
\]
This means that to identify which vertices are in $B_i$ for a given $i$, 
one should iterate over the vector $t$, which contains tentative distances, and select those vertices whose tentative distance are in the correct range. Again this is a filtering operation where the filter we want to apply on the vector $t$ is 
\[
i\Delta \leq t < (i+1)\Delta,
\]
for a given $i$. Applying this filter on the vector $t$ returns a binary vector $t_{B_i}$, where
\[
t_{B_i} = (i\Delta \leq t < (i+1)\Delta),
\]
and the locations of the 1s in $t_{B_i}$ are the vertices that are in the current bucket $B_i$.

\subsection{Edge relaxation}
The bulk of the computation for the $\Delta$-stepping algorithm is in the relaxation of light and heavy edges. Recall that in this phase of the algorithm, all light (or heavy) edges are relaxed simultaneously to create
a set of requests, where each request identifies the reachable vertex and the new tentative distance from a particular vertex in the current bucket.

Consider light edges from an arbitrary vertex $v$. The matrix $A_L$ contains all light edges, and the row corresponding to vertex $v$ (i.e. $a_v$) contains the light edges with $v$ as their source vertex. This means that the row corresponding to vertex $v$ also indicates the reachable vertices from $v$.

Notice that the $a_v$ also contains the weights of the light edges from $v$ (i.e., $\delta(v, v_j)$). In order to compute the new tentative distance to a reachable vertex $v_j$, the following operation
\[
new\_dist_j = tent(v) + \delta(v, v_j)
\]
has to be applied to the edge $(v, v_j)$.  To compute the tentative distance to all reachable nodes, $tent(v)$ has to be added to all elements in $a_v$.

In the language of linear algebra, this operation on $a_v$ is similar to a scaled vector addition or AXPY 
operation, i.e.,
\[
y = ax + y,
\]
with two key differences. First, the ($\min, +$) semiring is used instead of the ($+, \times$) semiring. Second, the result is written into a new vector instead of overwriting $a_v$. Therefore, computing the set of requests for a particular vector $v$ can be described as the following linear algebra operation
\[
\overbrace{Req_v}^y = \underbrace{t[v]}_a + \overbrace{a_v}^x,
\]
where $t$ is the tentative distance to all vertices. Notice that this is similar to the linear algebraic formulation of edge-centric operations, in which each operation is applied point-wise.

Generalizing to relaxing all light edges requires us to apply the above operation to all rows in $A_L$. This means the set of requests can be viewed as a $|V| \times |V|$ matrix $Req$ that has been partitioned such that
\[
Req \rightarrow 
\left(\begin{array}{c}
Req_0\\ \hline
\vdots \\ \hline 
Req_{|V|-1}
\end{array}
\right)
=
\left(\begin{array}{c}
t[0] + a_0\\ \hline
\vdots \\ \hline 
t[|V|-1] + a_{|V|-1}
\end{array}
\right),
\]
where each element in a column in $Req$ represent a new tentative distance for the corresponding vertex to that column.

Among all possible new tentative distances for a vertex, i.e., all values in the same column of $Req$, only the minimum of those values  could potentially be the new tentative distance. This means that a reduction operation with the $\min$ operator has to be applied on all elements in the same column. Applying the $\min$ operator to all columns returns a vector $t_{Req}$ of possibly new tentative distances for all vertices. This is also equivalent to
\[
t_{Req} = A_L^T t.
\]

However, only out-going edges from vertices in the current bucket $B_i$ participate in the edge relaxation phase. This means that a filtering operation has to be applied to the vector $t$ to filter out vertices that are not in the current bucket. Mathematically, this filtering can be described as
\[
t \circ t_{B_i},
\]
which means the vector of requests is computed via matrix-vector multiplication over the ($\min, +$) semiring by 
\[
t_{Req} = A_L^T (t \circ t_{B_i}).
\]

Finally, edge relaxation updates the membership of vertices in each bucket if the tentative distance for the vertices has been updated. This may reintroduce vertices back into bucket $B_i$. To identify vertices that are reintroduced, we note that these vertices must satisfy the following:
\begin{enumerate}
    \item Their tentative distances were updated in the current processing phase of relaxation, i.e. $(t_{Req} < t)$, and
    \item Their new tentative distance assigns them into bucket $B_i$, i.e.
    \[
t_{B_i} = (i\Delta \leq t_{Req} < (i+1)\Delta).
\]
\end{enumerate}
This means that vertices that are reintroduced into bucket $B_i$ can be identified with
\[
t_{B_i} = (i\Delta \leq t_{Req} < (i+1)\Delta) \circ (t_{Req} < t).
\]
%
After identifying the vertices that are reintroduced, we can finalize the 
new tentative distances for all vertices with
\[
t = \min(t, t_{Req}).
\]

\subsection{Tracking processed vertices}
To ensure that the heavy edge relaxation can proceed with the appropriate heavy edge sets, it is necessary to keep track of which vertices were processed in the recently emptied bucket. From a set operation point of view, this is simply a union operation. Hence the set $S$ is updated by accumulating $t_{B_i}$ to $S$ as follows
\[
S = ((S + t_{B_i}) > 0).
\]

The final $\Delta$-stepping algorithm defined using the language of linear algebra is shown on the left side of Fig.~\ref{fig:DSSSP}.

\section{Implementations}
We implemented our linear algebraic $\Delta$-stepping algorithm using GraphBLAS APIs in both C and C++. For the C implementation we linked to the SuiteSparse library \cite{suitesparse}, while in our C++ implementation we used the GraphBLAS template library (GBTL)~\cite{gbtl}. As these implementations are sequential in nature, we additionally implemented a parallel version of the $\Delta$-stepping algorithm in C with OpenMP tasks~\cite{openmp}.


\subsection{GraphBLAS implementations with SuiteSparse and GBTL}

Implementation of the linear algebraic $\Delta$-stepping algorithm discussed in Sec.~\ref{sec:ladsssp} using the GraphBLAS C and C++ interfaces was relatively straightforward. The GBTL implementation is available online~\cite{gbtl_sssp} while Fig.~\ref{fig:gbcode} shows our implementation using SuiteSparse. To help identify the connections between algorithmic steps and their implementations, all the operations listed in the linear algebraic formulation in Fig.~\ref{fig:DSSSP} are also reported within comments in the code in Fig.~\ref{fig:gbcode}.
In this section, we introduce the main functions from the GraphBLAS interface that are required to implement our linear algebraic $\Delta$-stepping and highlight major caveats when using them in our implementation.

The first function of interest is 
\begin{center}
\footnotesize
\texttt{GrB\_apply(out, mask, accum, op, in, desc)},
\end{center}
which is used in the creation of filters. Specifically, the function
computes into \texttt{out}, the transformation of the values in \texttt{in} using a unary function \texttt{op}. Optionally, \texttt{mask} can be used to control which elements are actually stored in \texttt{out} while \texttt{accum} defines a binary operation to accumulate the result rather than assigning it. The \texttt{desc} parameters can be used to set optional flags 
for controlling further aspects of the operation, such as whether the output has to be cleared before storing the new result.
Both the vector and the matrix variants of the \texttt{GrB\_apply} function are used in the code for implementing filtering operations.
For instance, Fig.~\ref{fig:gbcode}, lines 20--21 show the implementation of the heavy edges selection $A_H = A \circ (A > \Delta)$. This operation requires two subsequent calls to \texttt{GrB\_apply}. The first call implements the filter $(A > \Delta)$, whose output is then used as a mask by the second call to compute the Hadamard product. 
Notice that as above, even if the algorithm only requires a filter operation (e.g., $t \geq i\Delta$), \texttt{GrB\_apply} has to be used twice to avoid storing elements where the filtering condition is falsified (typically false or zero values). For example, Fig.~\ref{fig:gbcode}, lines 27--28 implements the filtering operation above. The first call to \texttt{GrB\_apply} creates the filtered output and the second call uses it as a mask to store only true values.

The next GraphBLAS function used to implement our algorithm is
\begin{center}
\footnotesize
    \texttt{GrB\_eWiseAdd(out, mask, accum, op, in1, in2, desc)},
\end{center} which applies either a monoid, semiring, or binary operation \texttt{op}
to \texttt{in1} and \texttt{in2}, element-wise. `\texttt{Add}' in the name refers to operation on the union of elements from each input vector. Therefore, \texttt{GrB\_eWiseAdd} applies to pairs of input elements where at least one is defined; if both values are defined, the operation returns the result; otherwise the single defined value is used. The parameters \texttt{mask}, \texttt{accum}, and \texttt{desc} have the same meaning as in \texttt{GrB\_apply}.
In our context, in addition to performing element-wise operations such as computing the minimum of two vectors, we also use \texttt{GrB\_eWiseAdd} to perform filtering when the condition is between two vectors. For example, in Fig.~\ref{fig:gbcode}, line 48 \texttt{GrB\_eWiseAdd} is used to compute the
condition $(t_{Req} < t)$.

The last GraphBLAS function we use is \texttt{GrB\_vxm(out, mask, accum, op, v, M, desc)} which computes the (row) vector-matrix multiplication between vector \texttt{v} and matrix
\texttt{M} on the semiring \texttt{op} and stores the resulting vector in \texttt{out}. Also in this case, the parameters \texttt{mask}, \texttt{accum}, and \texttt{desc} described above can be optionally used. In our implementation, we use the function as shown in Fig.~\ref{fig:gbcode}, lines 43 and 60, to compute the
relaxation of light and heavy edges using the vector-matrix operation on the ($\min, +$) semiring. 


\begin{figure*}
{\scriptsize
\begin{lstlisting}[language=C,numbers=left]
GrB_Info sssp_delta_step(const GrB_Matrix A, type_t d, GrB_Index src, GrB_Vector ** paths){
    // Global scalars:
    delta=d;
    // Define operators, scalar, vectors, and matrices
    ...

    // t[src] = 0
    GrB_Vector_setElement(t, 0, src);

    // Create A_L and A_H based on delta:
    GrB_Matrix Ah, Al, Ab;
    GrB_Matrix_new(&Ah, GrB_FP64, n, m);
    ...

    // A_L = A .* (A .<= delta)
    GrB_apply(Ab, GrB_NULL, GrB_NULL, delta_leq, A, GrB_NULL );
    GrB_apply(Al, Ab, GrB_NULL, GrB_IDENTITY_FP64, A, GrB_NULL );

    // A_H = A .* (A .> delta)
    GrB_apply(Ab, GrB_NULL, GrB_NULL, delta_gt, A, GrB_NULL );
    GrB_apply(Ah, Ab, GrB_NULL, GrB_IDENTITY_FP64, A, GrB_NULL );

    // init i = 0
    i_global = 0;

    // Outer loop: while (t .>= i*delta) != 0 do
    GrB_Vector_apply(tgeq, GrB_NULL, GrB_NULL, delta_i_geq, t, GrB_NULL);
    GrB_Vector_apply(tcomp, tgeq, GrB_NULL, GrB_IDENTITY_BOOL, t, GrB_NULL);
    GrB_Vector_nvals(&t_comp_size, tcomp);
    while (t_comp_size > 0) {
        //s=0
        GrB_Vector_clear(s);

        // tBi = (i*delta .<= t .< (i+1)delta)
        GrB_Vector_apply(tB, GrB_NULL, GrB_NULL, delta_i_range, t, clear_desc);
        // t .* tBi
        GrB_Vector_apply(tmasked, tB, GrB_NULL, GrB_IDENTITY_FP64, t, clear_desc);
    
        // Inner loop: while tBi != 0 do
        GrB_Vector_nvals(&tm_size, tmasked);
        while (tm_size > 0) {
            // tReq = A_L' (min.+) (t .* tBi)
            GrB_vxm(tReq, GrB_NULL, GrB_NULL, min_plus_sring, tmasked, Al, clear_desc);
            // s = s + tBi
            GrB_eWiseAdd(s, GrB_NULL, GrB_NULL, GrB_LOR, s, tB, GrB_NULL);

            // tBi = (i*delta .<= tReq .< (i+1)*delta) .* (tReq .< t)
            GrB_eWiseAdd(tless, tReq, GrB_NULL, GrB_LT_FP64, tReq, t, clear_desc);
            GrB_Vector_apply(tB, tless, GrB_NULL, delta_i_range, tReq, clear_desc);

            // t = min(t, tReq)
            GrB_eWiseAdd(t, GrB_NULL, GrB_NULL, GrB_MIN_FP64, t, tReq, GrB_NULL);

            GrB_Vector_apply(tmasked, tB, GrB_NULL, GrB_IDENTITY_FP64, t, clear_desc);
            GrB_Vector_nvals(&tm_size, tmasked);

        }
        // tReq = A_H' (min.+) (t .* s)
        GrB_Vector_apply(tmasked, s, GrB_NULL, GrB_IDENTITY_FP64, t, clear_desc);
        GrB_vxm(tReq, GrB_NULL, GrB_NULL, min_plus_sring, tmasked, Ah, clear_desc);

        // t = min(t, tReq)
        GrB_eWiseAdd(t, GrB_NULL, GrB_NULL, GrB_MIN_FP64, t, tReq, GrB_NULL);
        
        // i=i+1
        i_global++;
        GrB_apply(tgeq, GrB_NULL, GrB_NULL, delta_i_geq, t, clear_desc);
        GrB_apply(tcomp, tgeq, GrB_NULL, GrB_IDENTITY_BOOL, t, clear_desc);
        GrB_Vector_nvals(&t_comp_size, tcomp);
    }

    // Set the return paths
    (*paths) = &t;
    return (GrB_SUCCESS);
}
\end{lstlisting}
}
\caption{SuiteSparse implementation of the linear algebraic algorithms in Fig.~\ref{fig:DSSSP}.}
\label{fig:gbcode}
\end{figure*}

\subsection{Implementing Filters}
The main implementation hurdle we encountered was the creation of filters for performing the necessary selection of vertices and edges during the course of the algorithm. 

We illustrate a peculiarity in creating a filter within GraphBLAS. Recall that one of the filters we required was the filter 
\[
t_{Req} < t.
\]
One option of creating this filter is to use the \texttt{eWiseAdd} function which applies a binary operator on its two input vectors to create a third vector containing the eventual filter. An intuitive understanding of $t_{Req} < t$ is that the output should be a binary vector in which elements are 1 when the inequality evaluates to true, and 0 otherwise. However, intuition fails us here. When both input values exist, then the operation behaves as expected, i.e. returning a 1 (true) or a 0 (false). However, when one of the input values is not present, the output of \texttt{eWiseAdd} is the single value that is present. This means that if a value in $t$ was present and no new requests update the tentative distance for that particular vertex, the check will return the value of $t$, which will evaluate to 1 (true), instead of the expected 0 (false). 

The reason for this behavior is that while the specification of \texttt{eWiseAdd} allows for the  operator to be a binary operator that is not commutative, the behavior specified for \texttt{eWiseAdd} assumes that the operator is commutative. This discrepancy could potentially be a stumbling block to an uninformed developer.  

A software solution to this unintuitive behavior is to apply $t_{Req}$ as an output mask to the call to $\texttt{eWiseAdd}$. Since we are only interested in values that exist in $t_{Req}$, $t_{Req}$ can be a mask to prevent unwanted values from being stored. Note that this solution works because $t_{Req}$ is never zero. If the value in $t_{Req}$ evaluates to zero and is stored, then the mask will be incorrect.

An alternative to using $\texttt{eWiseAdd}$ for this particular solution is to use $\texttt{eWiseMult}$. However, because this intersects the index sets of the two inputs, this solution would not work if one requires the filtering operation to allow new values (e.g. adding to a set). For example, assume that a tentative distance has been introduced to $t_{Req}$ that was not present in $t$, the vector of tentative distances. In this case, the inequality applied using $\texttt{eWiseMult}$ will incorrectly return no value (false); however, as undefined values of $t$ should default to $\infty$, the correct value should be true.

 

\section{Performance Results}
Our focus in this paper is to demonstrate a systematic approach for translating vertex- and edge-centric graph algorithms to the language of linear algebra. Nonetheless, in this section we show the performance of the different GraphBLAS implementations and highlight possible performance improvements for future GraphBLAS library implementations.
Since GBTL is a reference implementation for the GraphBLAS and has not been designed with performance in mind,  we do not report performance numbers in this case.   
\subsection{Experimental Setup}
We used two systems in the testing our implementations of delta-stepped SSSP. We ran both our sequential SuiteSparse implementation and our direct C implementation on a system with an Intel Xeon E5-2667 v3 CPU running at a fixed frequency of 3.20GHz. For mutithreaded scaling experiments, we ran on a quad-core Intel i7-7700K CPU running at a fixed frequency of 4.20GHz. Our sequential implementations were compiled with GCC version 7.2.1 on the Xeon machine and our parallel code was compiled with OpenMP and GCC version 6.4.1 on the i7 machine. Timing on each platform was performed with calls to the Intel RDTSC assembly instruction. 

For all tests, we used real-world graphs collected by the Stanford Network Analytics Platform (SNAP) \cite{snapnets} and the GraphChallenge \cite{graphchallenge}. 
These input data are symmetric and undirected graphs with unit edge weights; although our implementations can also operate on directed, weighted graphs. We ran our implementations with the input parameter $\Delta=1$.


\subsection{Opportunities for Fusion}
Note that most of the function calls in our implementations are related to implementing filtering operations. As filtering operations are point-wise operations, they are inherently memory-bounded. By performing loop fusion on these function calls, the overhead of the function calls and excessive data movement can be reduced. 

We demonstrate these potential savings with our C implementation, where the following fusion opportunities were exploited:
\begin{enumerate}
    \item Hadamard product and vector matrix multiplication i.e.
    \[
    t_{Req} = A_L^T(t \circ t_{B_i})
    \]
    were fused together into a single operation.
    \item The three vector operations required to compute $t_{B_i}$, $s$, and $t$ are fused together as the computation of one of the three outputs is dependent on at least one of the two vectors.
\end{enumerate}

\begin{figure*}
    \centering
    \includegraphics[width=1.0\textwidth]{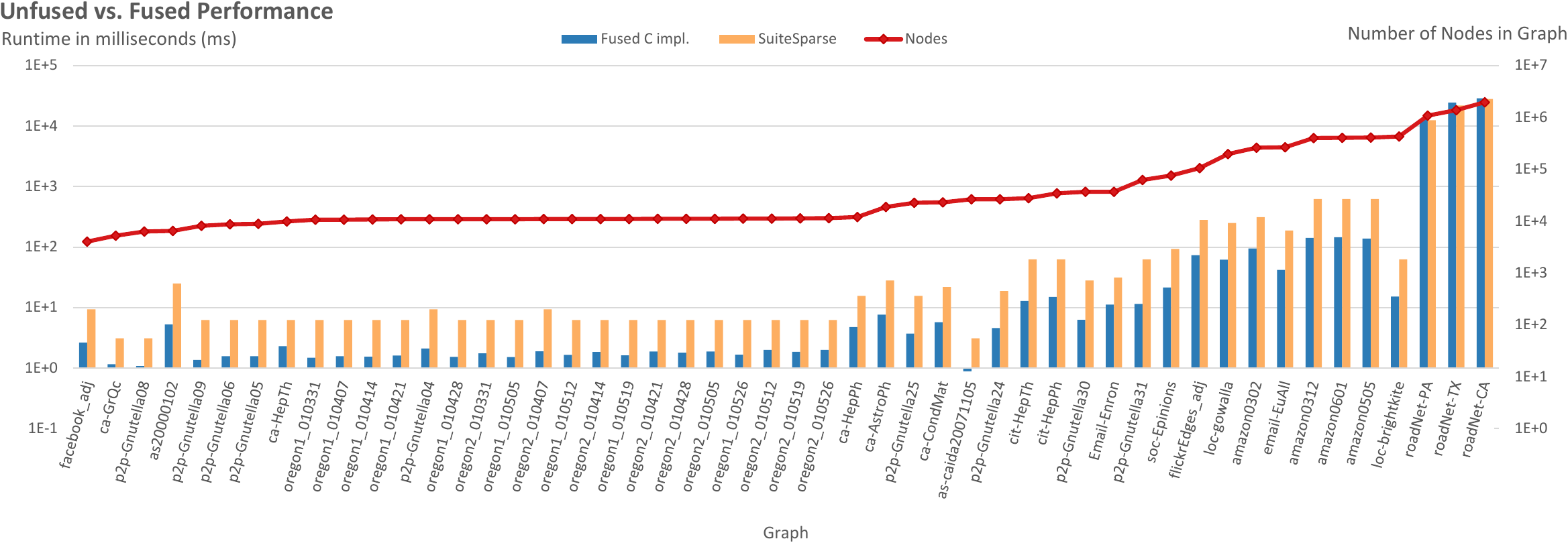}
    \caption{On average a 3.7x improvement in performance is attained by our sequential C implementation over SuiteSparse over a range of graphs by fusing operations. Graphs are sorted by ascending number of nodes, which is plotted on the second vertical axis.}
    \label{fig:SS_vs_fused}
\end{figure*}

The importance of identifying opportunities for fusion is evident from Fig.~\ref{fig:SS_vs_fused}, which compares the performance of an un-fused implementation (i.e. SuiteSparse) with one with fusion and elision of function calls via preprocessor-defined macros. This approach yielded, on average, a 3.7 times improvement over the functionally equivalent GraphBLAS implementation in SuiteSparse.  

\subsection{OpenMP Task Based Parallelism}
Given that both GBTL and SuiteSparse are sequential implementations, we made an initial attempt to parallelize our C implementation with OpenMP’s task parallelism pragmas to see possible benefits of parallelism.
Unlike the parallelism scheme proposed by Meyers and Sanders, we parallelized the vector and filtering operations in the fused sequential implementation. Specifically, the creation of the light and heavy edges are independent and were each made into a task. The computation and filtering of vectors was performed by splitting the vector into evenly-sized tasks.

\begin{figure*}
    \centering
    \includegraphics[width=1.0\textwidth]{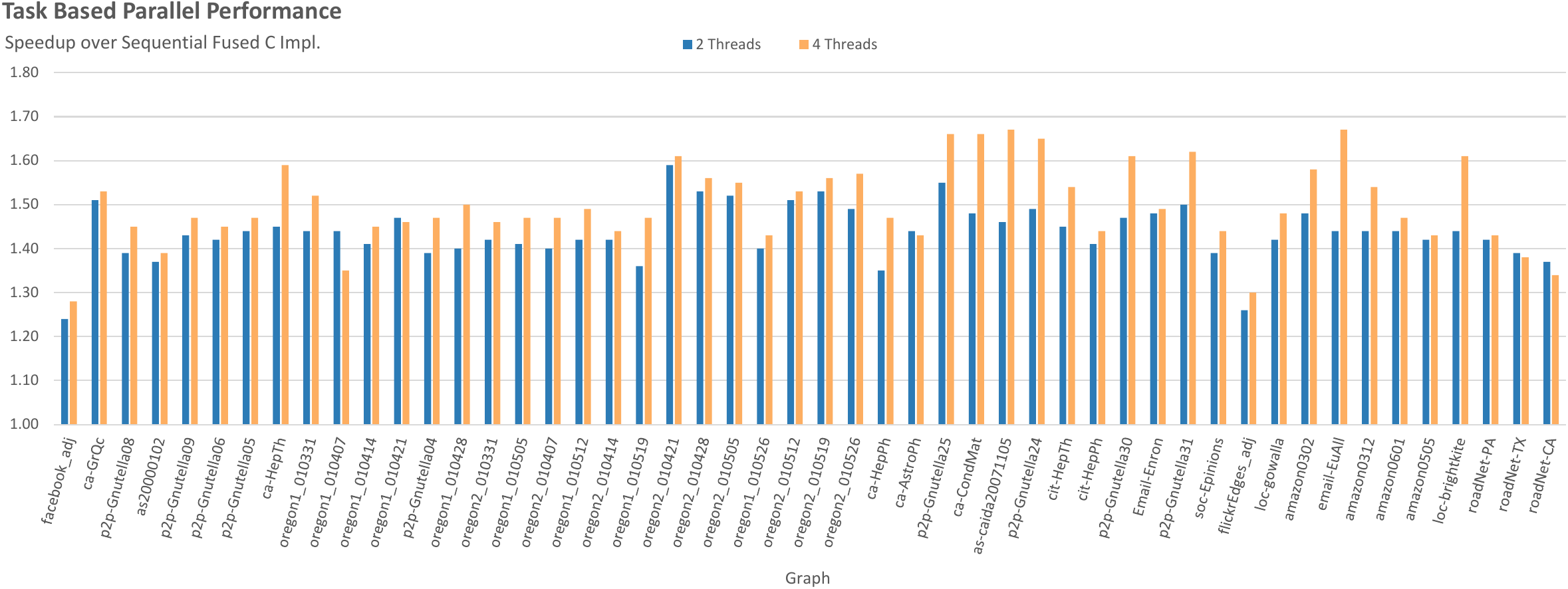}
    \caption{Performance of the $\Delta$-stepping C implementation on 2 and 4 threads, normalized to sequential performance. Note that the graphs are sorted along the x-axis in ascending number of nodes.}
    \label{fig:SSSP_par}
\end{figure*}

Fig.~\ref{fig:SSSP_par} shows the performance of our direct linear algebra to C approach as parallelized using OpenMP tasks. The performance of each increasing number of threads is normalized to the fused sequential C implementation shown in Fig.~\ref{fig:SS_vs_fused}. The overall results show an average of 1.44x and 1.5x speedup for running with two and four threads, respectively.

We note that the performance benefits of multithreading our C implementation are limited by how much we divide operations into tasks. In particular, the matrix filtering operations on $A_H$ and $A_L$ were noted to consume 35-40\% of the run time of the sequential implementation. Because each matrix is allocated to a single task, benefits of using more than two threads do not extend to these costly operations. Parallelizing within the matrix-vector operations and splitting the filtering operations for $A_H$ and $A_L$ into smaller tasks would allow more threads to participate in the filtering operation, thereby improving performance and scalability.


\section{Discussion}
Notice that the delta-stepping single source shortest path algorithm is both vertex-centric and edge-centric in nature. The initialization of the algorithm is vertex-centric in that the initialization operations are defined as operations on all vertices. However, the relaxation operation that updates the tentative distance is an edge-centric operation in that the edge relaxation operation is applied on light (or heavy) edges.  We believe that the linear algebraic formulation is sufficiently expressive to describe both kinds of algorithms. 

Another interesting point for discussion is that implementations of the original Dijkstra algorithm often requires the use of a priority queue so that the vertex with the minimum tentative distance from the source can be processed in each iteration. However, Meyers and Sanders showed that by setting the threshold $\Delta = 1$, the $\Delta$-stepping algorithm is analogous to the original  Dijkstra's algorithm. One possibility is for other algorithms that require the use of a priority queue to use a similar bucket approach. The assumptions and conditions under which this technique can be used have yet to be defined.

\section{Conclusion}
In this paper, we presented an initial attempt at a systematic approach of translating algorithms that have been described in terms of operations on edges and vertices to the language of linear algebra. Specifically, our approach is based on the identification of common vertex- and edge-based design patterns found in graph algorithms and their translation into linear algebraic forms. We demonstrated our approach by applying it to the $\Delta$-stepping algorithm to obtain its GraphBLAS implementations.
We also discussed implementation highlights required using both the C and C++ GraphBLAS APIs,
including possible improvements to the implementations of the underlying API libraries. 

We believe that the initial success of the OpenMP task parallelism suggest that much more can be achieved. Specifically, we believe that an approach to using OpenMP, as has been demonstrated by frameworks such as SuperMatrix~\cite{SuperMatrix:PPoPP08} and Magma~\cite{MAGMA}, can be used within the context of GraphBLAS to achieve better parallelism and to identify opportunities for operation fusion.

\section*{Acknowledgement}
The authors thank Marcin Zalewski of Northwest Institute for Advance Computing (NIAC) at Pacific Northwest National Laboratory for his valuable feedback.

This work was sponsored partly by the DARPA BRASS program under agreement FA8750-16-2-003, and by the NSF through award ACI 1550486. This material is based upon work funded and supported by the Department of Defense under Contract No. FA8702-15-D-0002 with Carnegie Mellon University for the operation of the Software Engineering Institute, a federally funded research and development center. [DM19-0235]

The content, views and conclusions presented in this document are those of the authors and do not necessarily reflect the position or the policy of the sponsoring agencies.

[DISTRIBUTION STATEMENT A] This material has been approved for public release and unlimited distribution.

\bibliographystyle{IEEEtran}
\bibliography{references,biblio,biblio2}


\end{document}